# Billions uncounted? Perspectives on rural population counts from gridded population datasets

Till Koebe[1], Emmanuel Letouzé[2], Tuba Bircan[3], Édith Darin[4], Douglas R. Leasure[4], Valentina Rotondi[4,5]

[1] Department of Computer Science, Saarland University
[2] Data-Pop Alliance
[3] Department of Sociology, Vrije Universiteit Brussel
[4] Demographic Science Unit, Nuffield Department of Population Health, University of Oxford
[5] Department of Business Economics, Health and Social Affairs, SUPSI

The paper titled "Global gridded population datasets systematically underrepresent rural population" by Josias Làng-Ritter et al. provides a valuable contribution to the discourse on the accuracy of global population datasets, particularly in rural areas. We recognize the importance of the work, as it adds significant validation efforts and draws further attention to rural population estimation challenges.This focus is valuable for improving datasets and fostering discussions on uncertainty communication. However, we caution on the conclusion drawn in the paper that the result "implies that rural population is, even in the most accurate dataset, underestimated by half compared to reported figures". Taking this statement seriously, it would constitute an undercounting of the world's population by 1-2.6 billion (by applying the study's reported bias of -80% to -32% for 2010 to the 2010 rural population of 3.24 billion according to the World Bank[1]). What we claim, here, is that researchers should exert greater caution in their claims to avoid ambiguity and non-academic press coverage with titles such as, "There could be billions more people on Earth than previously thought"[9]. We caution against such bold claims, if not properly backed by evidence, not only because of the methodological limitations outlined below, but also due to their potential to fuel harmful socio-political and environmental narratives as they could amplify climate anxiety, xenophobic or alarmist discourses and skepticism toward science and expert institutions.

We argue that the reported bias figures are less caused by actual undercounting of rural populations, but more so by contestable methodological decisions and the historic misallocation of (gridded) population estimates on the local level. Furthermore, we are emphasising that it is not possible to draw from the reported bias a coherent, global picture of the undercount of rural populations. We would like to highlight several critical points regarding the methodology and conclusions drawn in the study:

1. **Temporal misalignment between population estimates and resettlement data.** The study compares gridded population estimates from approximately ten years before dam completion with reported resettlement figures. However, project inception, usually entailing land acquisition combined with resettlement incentives often begin much earlier—sometimes decades before dam completion. Using two of the three German dam projects included in the study as examples, it took almost 30 years from inception to completion for the Brombachsee project and the settlements in the flooded area were purchased soon after project inception[8]. Thus, ten years before completion, around the time of the reference year, the affected population has left, while still being counted as resettled, which would inflate the bias estimates.

2. **Percentage-based metrics are sensitive to small denominators.**
   In the case of the second German dam, the Rothsee dam, roughly 13 people still lived in the area at the time of resettlement[7]. While a gridded population estimate of e.g. 7 or 8 for that area is not necessarily a "big miss", a percentage error (such as the sMAPE used in the study) looks large and may not be appropriate for extrapolating to country-level population numbers. Furthermore, extrapolating from three dam projects (affecting roughly 100 people) built on a methodology that does not properly seem fit-for-purpose to state that over 50% of the German rural population goes uncounted as alluded to in Fig. 7 and Fig. 8 of the study might risk overgeneralization, which also lacks backing from other data collection efforts such as recent census rounds in Germany. We expect that both the temporal misalignment and the metric argument is not limited to the German example, but holds true for other projects under study as well.

3. **Unreliable linkage between resettlement data and spatial extent.** To assess gridded population counts it is key to have a reliable link between population count and spatial extent of the enumerated location. The authors noticed that the polygons they used, which are from the GeoDar dataset, have a systematically smaller surface than the reservoir area reported by ICOLD. In order to circumvent the issue, the authors apply a fixed factor of 1.23 (that is an additional 23%) to the population number extracted from the gridded population. This procedure questions the reliability of the ground truth regarding fine-scale population counts.

4. **Lack of detailed settlement maps before 2010.**
   The study acknowledges, and we further emphasize it, that historical gridded population data are inherently limited due to the lack of detailed settlement maps before 2010. As described in the study, the bias in population counts decreased from -80% to -32% over the period of 2000 to 2010, which contradicts the study's claim of a consistent underrepresentation of the rural population. Since gridded population layers in this study usually disaggregate census counts with the help of satellite imagery-derived pixel-level maps of human presence, inaccurate spatial patterns of detected human presence are, in our view, an important source of potential bias in hyper-localized population estimates. Typically, populations in small rural settlements are assigned to the nearest larger settlements, which does not affect the size of the rural population per se, just its allocation. Of the data sources evaluated, the GHS-POP dataset, which had the most severe reported bias, is also the most sensitive to this fine-scale spatial misallocation because it uses satellite-based settlement maps to restrict where populations are mapped. This fine-scale spatial misallocation does not necessarily indicate overall bias in population estimates across larger areas. We observed similar issues in detecting small rural settlements in Colombia, especially in the forested areas of the Amazonia[2]. Since 2010, much progress has been made in detecting objects from satellite imagery[5,6] and we urge global gridded population projects to take recent advancements in this area into account.

5. **Lack of digitization of census techniques before 2010**.
   The authors speculate that the underrepresentation of rural populations may be due to rural undercounts in censuses. This claim lacks proper backing as their data validation approach does not allow them to disentangle the effects of the different components in the data processing pipelines. Also, this would contradict quality assurance measures of census efforts around the world. The example of Paraguay's census which may have missed a quarter of the population does not relate to its rural population exclusively. The Colombian National Statistics Office DANE, for example, published post-census survey results for its 2018 census that quantifies coverage issues[3]. And here again, technological advancements in census operations, from providing digital maps to surveyors and tracking their GPS to ensure that all settlements are covered, to online census forms, let us assume that miscounting is less of an issue today[4].

6. **Non-representative sample of rural areas.**
   The 2000s were a significant period for dam-related resettlement in China, such as the Three Gorges Dam, the Wanjiazhai Dam, the Jiangya Dam and the Jinpen Dam, to name a few, which led to resettlements of millions of people[11]. This context is critical, as China accounts for over two-thirds of the study's sample in the reference year 2000 (203 out of 307 areas). While the authors conduct a robustness check excluding China and report similar aggregate results, the performance of individual datasets shifts markedly: GRUMP's bias improves from -66.9% to -12.2%, while WorldPop's bias worsens from -53.4% to -71.2%. This sensitivity calls into question the stability of the paper's conclusions regarding dataset performance, especially given their role in guiding recommendations for rural data use.

Beyond the points outlined above, we would like to emphasize that the paper misses out on providing arguments that the dams studied with its natural proximity to water and contingent vegetation provide a representative picture of the rural population in the respective countries. While rural undercounting in censuses is a known challenge in the official statistics community[10], the authors fall short of exploring other potentially important explanations for the biases observed. For example, dam watersheds may not be representative of all rural populations due to their distinct features such as the proximity to water bodies, steep terrain, hydrophile vegetation and corresponding downstream socio-economic characteristics such as the absence/presence of certain industries (e.g. fisheries, water-intensive industries etc.). Furthermore, as some gridded population datasets such as WorldPop take these spatial features into account in their modelling approach, their measured model performance at dams specifically may not represent their overall model performance in rural areas as some predictors may constantly exhibit extreme values (i.e. distances to water bodies near zero) with little variation.

# Conclusion

We recognize the efforts put into this research and appreciate its contribution to the field. However, we feel that key claims in the study are overly bold, not properly backed by evidence and lack a cautious and nuanced discussion. We hope these points will be taken into account in future discussions and refinements of population estimation methodologies.

# Competing interests statement

The authors declare no competing interests.

# Author contributions statement

DL, ED, TK and VR wrote the initial draft of the manuscript. DL, ED, EL, TB, TK and VR revised the initial draft.